\documentclass[aps,prb,twocolumn,floats,amsmath,amssymb,superscriptaddress]{revtex4-2}
\usepackage[utf8]{inputenc}
\usepackage{graphicx}
\usepackage{epsfig}
\usepackage{amsfonts}
\usepackage{amsmath}
\usepackage{natbib}
\usepackage{multirow}
\usepackage{bm}
\usepackage{epstopdf}
\usepackage{color}
\usepackage{hyperref}
\usepackage{mathtools}
\usepackage{cases}
\usepackage{physics}
\usepackage{amssymb}
\usepackage[noabbrev]{cleveref}
\usepackage{longtable}
\usepackage{booktabs} 
\usepackage{xspace}
\newcommand{\ie}{{\it i.e.}, }

\newcommand{\exciting}{{\usefont{T1}{lmtt}{b}{n}exciting\xspace}}
\Crefname{equation}{Eq.}{Eqs.}

\usepackage{xcolor}
\begin{document}

\title{An efficient approach to electron-phonon coupling calculations of heterostructures
within an all-electron framework}

\author{Ignacio Gonzalez Oliva}
\thanks{New address of I. Gonzalez Oliva: Department of Physics, University of Central Florida, Orlando, Florida 32816, United States }
\affiliation{Physics Department and CSMB, Humboldt-Universit\"{a}t zu Berlin, zum Großen Windkanal 2, 12489 Berlin, Germany}

\author{Sebastian Tillack}
\affiliation{Physics Department and CSMB, Humboldt-Universit\"{a}t zu Berlin, zum Großen Windkanal 2, 12489 Berlin, Germany}

\author{Fabio Caruso}
\affiliation{Institut f\"{u}r Theoretische Physik und Astrophysik, Christian-Albrechts-Universit\"{a}t zu Kiel, D-24098 Kiel, Germany.}
\affiliation{Institut f\"{u}r Physik, Universit\"{a}t Rostock, Rostock, Germany}

\author{Pasquale Pavone}
\affiliation{Physics Department and CSMB, Humboldt-Universit\"{a}t zu Berlin, zum Großen Windkanal 2, 12489 Berlin, Germany}

\author{Claudia Draxl}
\affiliation{Physics Department and CSMB, Humboldt-Universit\"{a}t zu Berlin, zum Großen Windkanal 2, 12489 Berlin, Germany}
\affiliation{European Theoretical Spectroscopic Facility (ETSF)}

\date{\today}
\begin{abstract}
A major challenge in calculating electron-phonon coupling at interfaces from first principles is the prohibitive computational cost due to the large supercells involved. Here, we present an efficient approach, which builds the electron-phonon matrix of weakly bound heterostructures from the properties of the isolated layers. The two key ingredients, the dynamical matrices and the potential response, are constructed in two steps. In the expansion step, these quantities are folded from the unit cells onto the supercell through an exact Fourier mapping. In the addition step, the interatomic force constants and the potential responses of the heterostructure are assembled blockwise from the corresponding quantities of the isolated layers, neglecting the weak interlayer contribution. The method is implemented in the electron-phonon module of the all-electron full-potential code \exciting, enabling efficient calculations of the electron-phonon self-energy. We demonstrate the approach for the phonon dispersion and spectral functions of bilayer WSe$_2$ and pyridine/MoS$_2$, in comparison with exact reference calculations.
\end{abstract}
\maketitle


\section{Introduction}
\label{sec:intro}

Stacking two-dimensional materials into weakly bound van der Waals (vdW) heterostructures has become a powerful route to tune electronic and optical properties beyond those of the individual layers~\cite{Geim2013,Novoselov2016}. In this regard, transition metal dichalcogenides (TMDCs) are prominent building blocks~\cite{Manzeli2017}, whether combined with one another into multilayered structures or interfaced with organic molecules to form hybrid inorganic organic systems (HIOS)~\cite{Xu2021,Ji2022,Benson2024,Cui2025}. These systems host rich phenomena, from interlayer and moiré excitons in multilayer TMDCs~\cite{Jiang2021} to interfacial charge transfer, hybrid and charge-transfer excitons, and strong electron-phonon coupling (EPC) in HIOS~\cite{Sarkar2017,Ulman2021,Bennecke2025,Chowdhury2025,GonzalezOliva2022,GonzalezOliva2026, Champagne2026}.

For the quantitative modeling of various excitations, many-body perturbation theory (MBPT) has become the standard for describing electron-electron and electron-hole interactions, through the $G_0W_0$ approximation for quasiparticle (QP) band structures and the Bethe-Salpeter equation (BSE) for optical excitations~\cite{GonzalezOliva2022,Champagne2026}, respectively. Likewise, a consistent treatment of self-energy effects from EPC within MBPT~\cite{Eiguren2008,Eiguren2009,Gonze2011,Ponce2014,Ponce2015,Marini2015,Giustino2017,Schebek2025} has become routine for semiconductors~\cite{Giustino2017} and more recently, has been demonstrated to be substantial also in HIOS~\cite{Eschmann2021,Eschmann2021_B,GonzalezOliva2026}. For complex systems, however, and in particular for interfaces, which typically require large supercells, the computational cost of MBPT renders its application often prohibitive. This not only applies to $GW$ and BSE, but also to the investigation of EPC effects, since one needs phonon calculations for the complete heterostructure. For the electronic and optical properties, this cost has motivated an expansion and addition screening (EAS) method for the non-interacting polarizability of vdW heterostructures~\cite{Xuan2019,Liu2019,Schebek2025_PRB}, a key ingredient in both $G_0W_0$ and BSE calculations. In this work, we extend this idea to EPC, an approach we refer to as EA-EPC. We have implemented it in the all-electron code \exciting~\cite{Gulans2014,RayaMoreno2026}, which is based on the (linearized) augmented planewave + local orbital (L)APW+LO method. The central quantity of this calculation is the EPC matrix, which depends on the dynamical matrix and the potential response upon the displacements of the nuclei. Our method constructs these ingredients for the heterostructure in the two steps that give it its name. In the expansion step, the dynamical matrix and potential response of each layer are mapped onto the supercell of the heterostructure. In the addition step, the contributions of the individual layers are combined into those of the full system. If the interaction between the layers is weak, this construction recovers the EPC of the heterostructure from calculations of its constituents alone, at a fraction of the computational cost of a full supercell treatment. 

We demonstrate the approach on two representative systems, bilayer WSe$_2$ and pyridine/MoS$_2$, comparing against exact reference calculations for the full heterostructure. For the TMDC, EA-EPC reproduces the spectral function and band-gap renormalization over a range of temperatures, for the HIOS, where the molecular vibrational modes are far more sensitive to the environment, it still captures the band gap renormalization, showing both the success and the limitations of the method. 

\section{Electron-phonon coupling}
\label{sec:epc}

The central quantities entering the electron self-energy due to EPC are the electron-phonon matrix elements~\cite{Giustino2017} corresponding to the first-order Kohn-Sham (KS) potential response $v^{(\mathbf{q}\alpha i)}_{\rm KS}(\mathbf{r})$ to a wave-like collective displacement with wavevector \textbf{q} of atom $\alpha$ along direction $i$,
\begin{equation}
  g_{mn,\nu}(\mathbf{k},\mathbf{q})
  = \sum_{\alpha,i}
    \frac{w_{\alpha i,\nu\mathbf{q}}}{\sqrt{2 M_\alpha\,\omega_{\nu\mathbf{q}}}}\,
    \langle \psi_{m\mathbf{k}+\mathbf{q}} |
    v^{(\mathbf{q}\alpha i)}_{\rm KS} |
    \psi_{n\mathbf{k}} \rangle \;,
  \label{eq:eph_matrix_elements}
\end{equation}
where $\psi_{n\mathbf{k}}$ are the KS states and $M_{\alpha}$ the atomic masses. The frequencies $\omega_{\nu\mathbf{q}}$ and eigenvectors $w_{\alpha i,\nu\mathbf{q}}$ of the phonons with wavevector $\mathbf{q}$ are obtained from the dynamical matrix $D_{\alpha i,\alpha' j}(\mathbf{q})$. The two ingredients that must be constructed for the heterostructure are (i) the dynamical matrix, which provides $\omega_{\nu\mathbf{q}}$ and $w_{\alpha i,\nu\mathbf{q}}$, and (ii) the potential response $v^{(\mathbf{q}\alpha i)}_{\rm KS}$. We consider the typical contributions to the electron self-energy due to EPC~\cite{Giustino2017}: The first is often called Fan-Migdal (FM) self-energy,
\begin{align}
& \Sigma^{\rm FM}_{nn'}(\omega,\mathbf{k},T)
  = \sum_{\nu,m} \int_{\rm BZ}  \frac{\mathrm{d}\mathbf{q}}{\Omega_{\rm BZ}}\,
    g^{\ast}_{mn,\nu}(\mathbf{k},\mathbf{q})\,g_{mn',\nu}(\mathbf{k},\mathbf{q}) \nonumber \\
    & \times \Bigg[  \frac{f_{m\mathbf{k}+\mathbf{q}}(T)+n_{\nu\mathbf{q}}(T)}
                  {\omega-\epsilon_{m\mathbf{k}+\mathbf{q}}+\omega_{\nu\mathbf{q}}+i\eta}
    +  \frac{1-f_{m\mathbf{k}+\mathbf{q}}(T)+n_{\nu\mathbf{q}}(T)}
             {\omega-\epsilon_{m\mathbf{k}+\mathbf{q}}-\omega_{\nu\mathbf{q}}+i\eta}
    \Bigg]\;,
  \label{eq:fm_self_energy}
\end{align}
where $f(T)$ [$n(T)$] are the temperature-dependent fermionic [bosonic] occupation numbers of the electrons [phonons], and $\eta$ is a smearing parameter. $\Omega_{\rm BZ}$ is the volume of the Brillouin zone (BZ).

The second, static contribution, is the Debye-Waller (DW) self-energy,
\begin{align}
  \Sigma^{\rm DW}_{nn'}(\mathbf{k},T)
  = \sum_{\nu}\int_{\rm BZ}\frac{\mathrm{d}\mathbf{q}}{\Omega_{\rm BZ}}\,
    g^{\rm DW}_{nn',\nu}(\mathbf{k},\mathbf{q})
    \left[ n_{\nu\mathbf{q}}(T)+\tfrac{1}{2} \right]\;,
  \label{eq:dw_self_energy}
\end{align}
with the DW matrix elements
\begin{align}
  g^{\rm DW}_{nn',\nu}(\mathbf{k},\mathbf{q})
  = \sum_{\substack{\alpha,i\\ \alpha',j}}
    \frac{w^{\ast}_{\alpha i,\nu\mathbf{q}}\,w_{\alpha' j,\nu\mathbf{q}}}
         {2\omega_{\nu\mathbf{q}}\sqrt{M_\alpha M_{\alpha'}}}\,
    \langle \psi_{n\mathbf{k}} |
    v^{(-\mathbf{q}\alpha i,\mathbf{q}\alpha' j)}_{\rm KS} |
    \psi_{n'\mathbf{k}} \rangle \;,
\label{eq:dw_matrix_elements}
\end{align}
which involve the second-order potential response $v^{(-\mathbf{q}\alpha i,\mathbf{q}\alpha' j)}_{\rm KS}$, evaluated within the rigid-ion approximation.

The spectral function is given by
\begin{widetext}
\begin{align}
  A_{n\mathbf{k}}(\omega,T)
  = -\frac{1}{\pi}\,
    \frac{{\rm Im}\,\Sigma_{nn}(\omega,\mathbf{k},T)}
         {\big[\omega-\epsilon_{n\mathbf{k}}-{\rm Re}\,\Sigma_{nn}(\omega,\mathbf{k},T)\big]^2
          +\big[{\rm Im}\,\Sigma_{nn}(\omega,\mathbf{k},T)\big]^2}\;,
\label{eq:spectral_function}
\end{align}
\end{widetext}
with $\Sigma_{nn}=\Sigma^{\rm FM}_{nn}+\Sigma^{\rm DW}_{nn}$. Both self-energy contributions are built from the same two ingredients, the dynamical matrix and the potential response, whose efficient construction for the heterostructure is the subject of the following section.

\section{The EA-EPC approach}
\label{sec:ea_epc}

\subsection{The expansion step}
\label{subsec:expansion}

In the expansion step, the dynamical matrix and the potential response are folded from the unit cell (uc) onto the supercell (sc) through Fourier mapping. For the dynamical matrix, $D(\mathbf{q})$, this yields,
\begin{align}
D^{\rm sc}_{\alpha^{\rm sc} i,\alpha'^{\rm sc} j}(\mathbf{q}^{\rm sc})
  = \frac{N^{\rm sc}_{\mathbf{q}}}{N^{\rm uc}_{\mathbf{q}}}
    \sum_{\mathbf{q}^{\rm uc}\in\mathcal{F}(\mathbf{q}^{\rm sc})}
    & e^{i\mathbf{q}^{\rm uc}\cdot(\mathbf{R}^{\rm uc}_{\alpha'}-\mathbf{R}^{\rm uc}_{\alpha})} \label{eq:expansion_dynmat} \\
    & \times\, D^{\rm uc}_{\alpha^{\rm uc} i,\alpha'^{\rm uc} j}(\mathbf{q}^{\rm uc}) \;, \nonumber
\end{align}
where $N^{\rm uc}_{\mathbf{q}}$ ($N^{\rm sc}_{\mathbf{q}}$) are the number of unit-cell (supercell) $\mathbf{q}$-points, $\mathbf{R}^{\rm uc}_{\alpha}$ the unit-cell lattice vector locating the supercell atom $\alpha^{\rm sc}$ relative to its unit-cell equivalent, $\boldsymbol{\tau}_{\alpha^{\rm sc}}=\boldsymbol{\tau}_{\alpha^{\rm uc}}+\mathbf{R}^{\rm uc}_{\alpha}$, and $\mathcal{F}(\mathbf{q}^{\rm sc})$ the set of unit-cell $\mathbf{q}$-vectors folding onto $\mathbf{q}^{\rm sc}$, \ie $\mathbf{q}^{\rm uc}=\mathbf{q}^{\rm sc}+\mathbf{G}^{\rm sc} (\mathbf{q}^{\rm uc})$, where $\mathbf{G}^{\rm sc} (\mathbf{q}^{\rm uc})$ is the supercell reciprocal-lattice vector that maps $\mathbf{q}^{\rm uc}$ back into the first supercell BZ . The potential response $v^{(\mathbf{q} \alpha i)}_{\rm KS}(\mathbf{r})$ follows an analogous relation,
\begin{align}
v^{{\rm sc},(\mathbf{q}^{\rm sc}\alpha^{\rm sc} i)}_{\rm KS}(\mathbf{r}) = 
\frac{N^{\rm sc}_{\mathbf{q}}}{N^{\rm uc}_{\mathbf{q}}} \sum_{\mathbf{q}^{\rm uc}\in\mathcal{F}(\mathbf{q}^{\rm sc})} e^{-i\mathbf{q}^{\rm uc}\cdot\mathbf{R}^{\rm uc}_{\alpha}}\, v^{{\rm uc},(\mathbf{q}^{\rm uc}\alpha^{\rm uc} i)}_{\rm KS}(\mathbf{r}) \;.
\label{eq:expansion_pot}
\end{align}
In the (L)APW+LO method, the unit cell is partitioned into muffin-tin (MT) spheres and the interstitial region. The potential response is represented differently in each region: inside the MT spheres as a spherical-harmonics expansion on a radial grid, and as a planewave expansion in the interstitial. The two regions therefore fold independently. For $\mathbf{r}$ in the MT sphere of atom $\beta^{\rm sc}$ ($\mathbf{r}=\boldsymbol{\tau}_{\beta^{\rm sc}}+\mathbf{s}$), this reads
\begin{align}
v^{{\rm sc},(\mathbf{q}^{\rm sc}\alpha^{\rm sc} i)}_{\rm KS}(\boldsymbol{\tau}_{\beta^{\rm sc}}+\mathbf{s})
  = \frac{N^{\rm sc}_{\mathbf{q}}}{N^{\rm uc}_{\mathbf{q}}} &
    \sum_{\mathbf{q}^{\rm uc}\in\mathcal{F}(\mathbf{q}^{\rm sc})}
    e^{-i\mathbf{q}^{\rm uc}\cdot(\mathbf{R}^{\rm uc}_{\alpha}-\mathbf{R}^{\rm uc}_{\beta})} \label{eq:expansion_pot_mt} \\
    & \times\, v^{{\rm uc},(\mathbf{q}^{\rm uc}\alpha^{\rm uc} i)}_{\rm KS}(\boldsymbol{\tau}_{\beta^{\rm uc}}+\mathbf{s}) \;, \nonumber 
\end{align}
while in the interstitial region the Fourier coefficients on the sc reciprocal lattice $\mathbf{G}^{\rm sc}$ are
\begin{align}
  \hat{v}^{{\rm sc},(\mathbf{q}^{\rm sc}\alpha^{\rm sc} i)}_{\rm KS}(\mathbf{G}^{\rm sc})
  = \frac{N^{\rm sc}_{\mathbf{q}}}{N^{\rm uc}_{\mathbf{q}}}
    & \sum_{\mathbf{q}^{\rm uc}\in\mathcal{F}(\mathbf{q}^{\rm sc})} 
    e^{-i\mathbf{q}^{\rm uc}\cdot\mathbf{R}^{\rm uc}_{\alpha}} \, \label{eq:expansion_pot_int}  \\
    & \times \hat{v}^{{\rm uc},(\mathbf{q}^{\rm uc}\alpha^{\rm uc} i)}_{\rm KS}(\mathbf{G}^{\rm sc}-\mathbf{G}^{\rm sc}(\mathbf{q}^{\rm uc})) \;. \nonumber
\end{align}
We note that the expansion is exact for commensurate \textbf{q}-grids and the supercell geometry consisting of ideal replica of the unit cell, which may not be strictly the case at interfaces. A full derivation is presented in Appendix~\ref{appendix:expansion}. 

\subsection{The addition step}
\label{subsec:addition}

In order to construct the interatomic force constants and potential response of the heterostructure (HS) from those of the individual layers, we neglect the interlayer contribution, assuming it to be weak. In this {\it addition step}, the interatomic force constants reduce to a block copy of the individual layer contributions as
\begin{align}
  \Phi^{\rm HS}_{\alpha i,\alpha' j}(\mathbf{R})
  = \sum_{\mu}
    \delta_{L(\alpha),\mu}\,\delta_{L(\alpha'),\mu}\,
    \Phi^{\mu}_{\alpha_\mu i,\alpha'_\mu j}(\mathbf{R})\;,
  \label{eq:addition_ifc}
\end{align}
where $L(\alpha)$ is the layer containing atom $\alpha$ and $\alpha_\mu$ its index within layer $\mu$. In other words, we set the force constants between atoms of different layers to zero (by ${\it layer}$ we mean the components of the interfacial systems). The potential response of the heterostructure is approximated by that of the layer containing the displaced atom, thereby neglecting the response of the other layer to that displacement,
\begin{align}
  v^{{\rm HS},(\mathbf{q}\alpha i)}_{\rm KS}(\mathbf{r})
  \approx \sum_{\mu} \delta_{L(\alpha),\mu}\, v^{\mu,(\mathbf{q}\alpha_{\mu} i)}_{{\rm KS}}(\mathbf{r})\;.
  \label{eq:addition_pot}
\end{align}
For $\mathbf{r}$ in the MT of an atom $\beta$ in the other layer, the response is transformed from the interstitial to the MT representation,
\begin{align}
v^{{\rm HS},(\mathbf{q}\alpha i)}_{\rm KS}(\boldsymbol{\tau}_\beta+\mathbf{s}) = & 4\pi \, e^{i(\mathbf{G}+\mathbf{q})\cdot\boldsymbol{\tau}_\beta} \sum_{l,m} i^{\,l} \sum_{\mathbf{G}}
    \hat{v}^{L(\alpha),(\mathbf{q}\alpha i)}_{{\rm KS}}(\mathbf{G})\, \nonumber \\
    & \times j_l(|\mathbf{G}+\mathbf{q}|\,s) \, Y^\ast_{lm}(\widehat{\mathbf{G}+\mathbf{q}})\,
    Y_{lm}(\hat{\mathbf{s}}) \;,
  \label{eq:addition_pot_mt}
\end{align}
with the spherical Bessel functions $j_l$  and the spherical harmonics $Y_{lm}$. In the interstitial, the cell-periodic parts are taken directly from the layer containing the displaced atom,
\begin{align}
\hat{v}^{{\rm HS},(\mathbf{q}\alpha i)}_{\rm KS}(\mathbf{G}) = \hat{v}^{L(\alpha), (\mathbf{q}\alpha i)}_{{\rm KS}}(\mathbf{G}) \;.
  \label{eq:addition_pot_int}
\end{align}
We note that in the addition step, we consider the individual layers and the heterostructure to have the same lattice parameters, the same \textbf{q}-grids, and the same maximum length of $|\mathbf{G}|$ for expanding the interstitial potential. The full derivations are given in the Appendix~\ref{appendix:addition}. The resulting dynamical matrix and potential response are used to compute the EPC matrix for the heterostructure and then fed into a Wannier-Fourier interpolation~\cite{Tillack2020,Tillack2025,Giustino2007,Eiguren2008} to produce the electron-phonon self-energies on fine \textbf{k}- and \textbf{q}-grids. 

\section{Computational details}
\label{subsec:comp_details}

We demonstrate our approach on two weakly bound systems: the WSe$_2$ bilayer and the HIOS pyridine/MoS$_2$. Both have previously served as examples for the EAS method in the context of $GW$ and BSE~\cite{Schebek2025_PRB}. The geometries and the computational parameters for the groundstate and $G_0W_0$ calculations are as reported in Ref.~\cite{Schebek2025_PRB}. The dynamical matrices and potential responses are computed within density-functional perturbation theory (DFPT)~\cite{Baroni2001,Tillack2026}. For WSe$_2$, we use a $6\times6\times1$ $\mathbf{q}$-grid for both the monolayer and the bilayer. For pyridine/MoS$_2$, the MoS$_2$ ingredients are obtained from a unit-cell calculation on a $6\times6\times1$ $\mathbf{q}$-grid and expanded to the $3\times3\times1$ supercell, while the pyridine ingredients are computed directly in the supercell on a $2\times2\times1$ $\mathbf{q}$-grid. The exact reference is the full DFPT calculation of the $3\times3\times1$ supercell on a $2\times2\times1$ $\mathbf{q}$-grid. For the calculation of the electron-phonon self-energies, the electron-phonon matrix elements are interpolated to a 100$\times$100$\times$1 $\mathbf{q}$-grid for WSe$_2$ and a 36$\times$36$\times$1 $\mathbf{q}$-grid for pyridine/MoS$_2$. The BZ integrals in Eqs.~\ref{eq:fm_self_energy} and~\ref{eq:dw_self_energy} are evaluated using tetrahedron integration, which allows for the calculation of the $\eta \rightarrow 0^{+}$ limit and ensures quick convergence. All timings reported below were obtained on a cluster at the Paderborn Center for Parallel Computing, using AMD EPYC 7763 64-core processors at a base frequency of 2.45 GHz.

\section{Results}
\label{sec:results}

The EA-EPC approach is applied only to the phonon-related quantities, that is, we use it to accelerate the construction of the dynamical matrices and the potential response. The electronic structure, by contrast, is treated exactly for the full heterostructure: the KS eigenvalues, wavefunctions, as well as the Wannier functions used in the interpolation, are all computed for the total system. 

\begin{figure}[t!]
\begin{center}
\includegraphics[width=0.95\columnwidth]{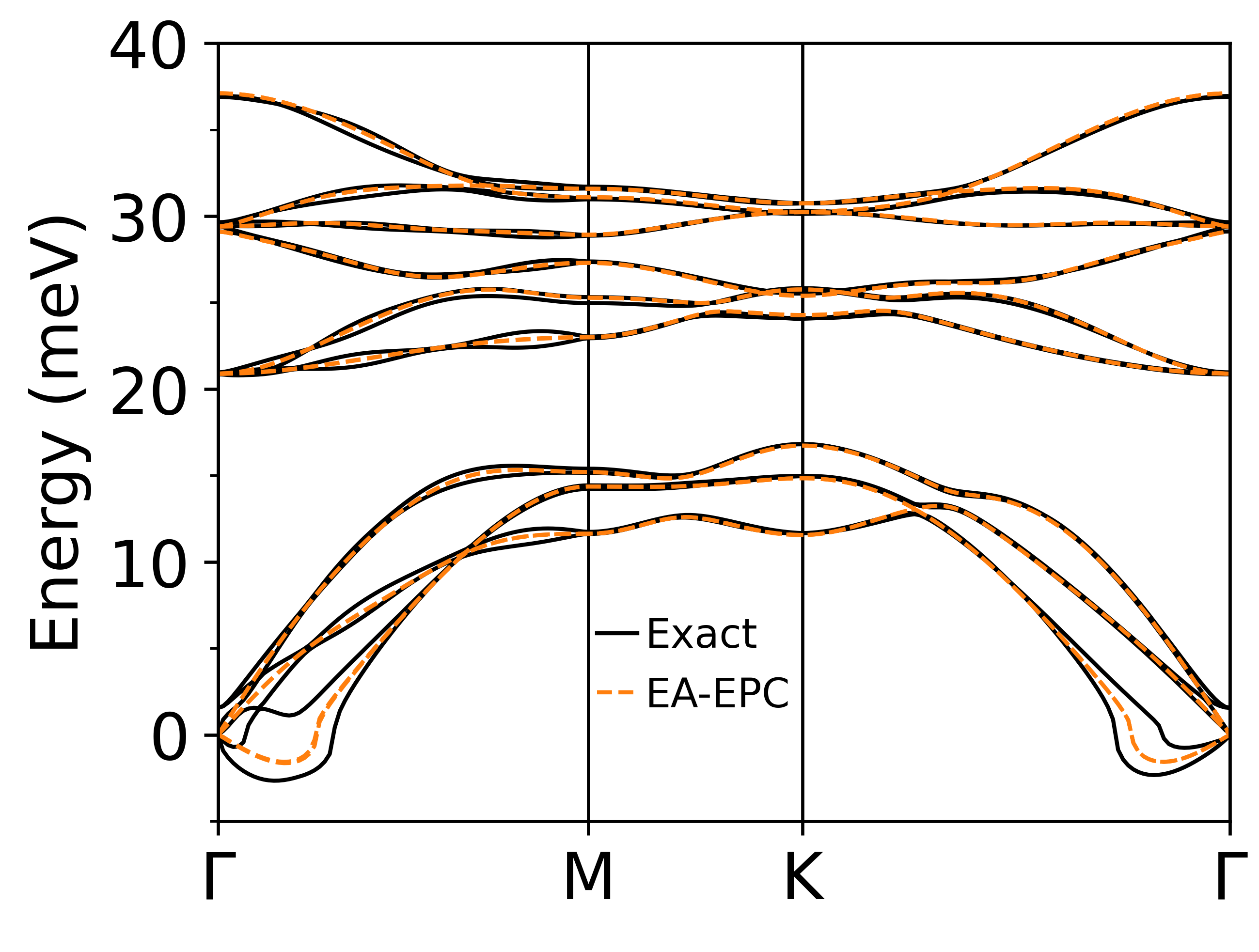}
\caption{Phonon dispersion of the WSe$_2$ bilayer obtained with the addition step of the EA-EPC approach (orange) and from the exact calculation (black). Since the interlayer force-constant blocks are set to zero, the addition step recovers the monolayer dispersion. 
\label{fig:phonon_dispersion_wse2_bilayer}}
\end{center}
\end{figure}

\subsection{WSe$_2$ bilayer}
\label{subsec:wse2}

We begin by assessing the addition step on the WSe$_2$ bilayer, whose phonon dispersion is shown in Fig.~\ref{fig:phonon_dispersion_wse2_bilayer}. Since the addition step neglects the force constants from the interface, the bilayer dispersion reduces to the superposition of two monolayers. The exact reference (black) lifts this degeneracy through the weak interlayer coupling, resulting in an overall very mild splitting of the monolayer phonon bands. Only for the acoustic modes, especially near $\Gamma$, the splitting becomes more pronounced. Across the rest of the path, the branches are essentially indistinguishable, confirming the dominance of the intralayer force constants.

\begin{figure}[t!]
\begin{center}
\includegraphics[width=0.95\columnwidth]{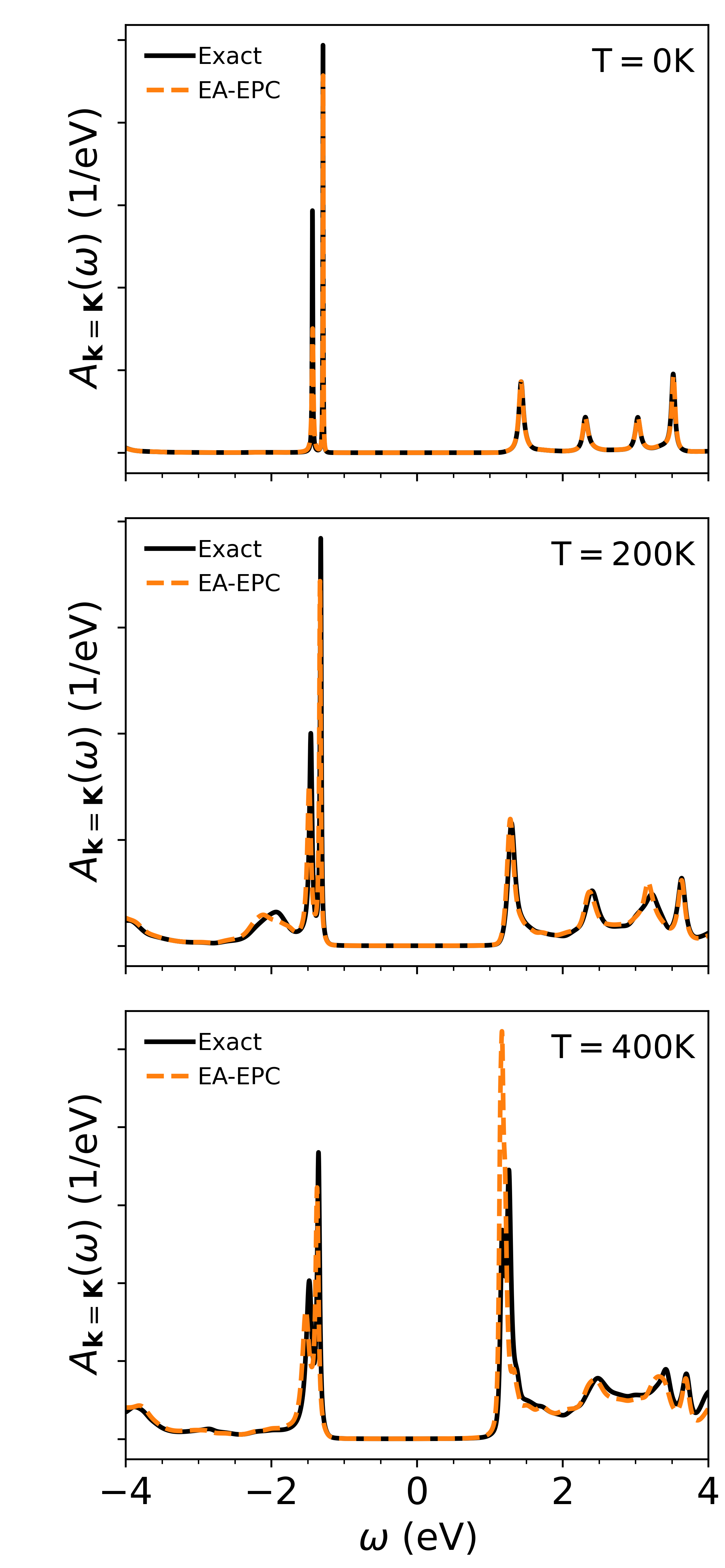}
\caption{Spectral functions of a WSe$_2$ bilayer at the K point of the first BZ obtained with the EA-EPC approach (orange) and the exact calculation (black) for temperatures of $T=0$K, $200$K, and $400$K.
\label{fig:spectral_function_wse2_bilayer}}
\end{center}
\end{figure}

We now assess the impact of this approximation on the electron spectral function $A_{n\mathbf{k}}(\omega)$, Eq.~\eqref{eq:spectral_function}, the observable that matters for spectroscopy. Figure~\ref{fig:spectral_function_wse2_bilayer} compares the spectral functions at the K point obtained by the two approaches for three representative temperatures. In all cases, the QP peak positions coincide almost perfectly. This agreement shows that omitting the interlayer coupling results in small enough changes that leave both the electron-phonon self-energy and spectral function essentially unchanged. 

\begin{figure}[t!]
\begin{center}
\includegraphics[width=0.95\columnwidth]{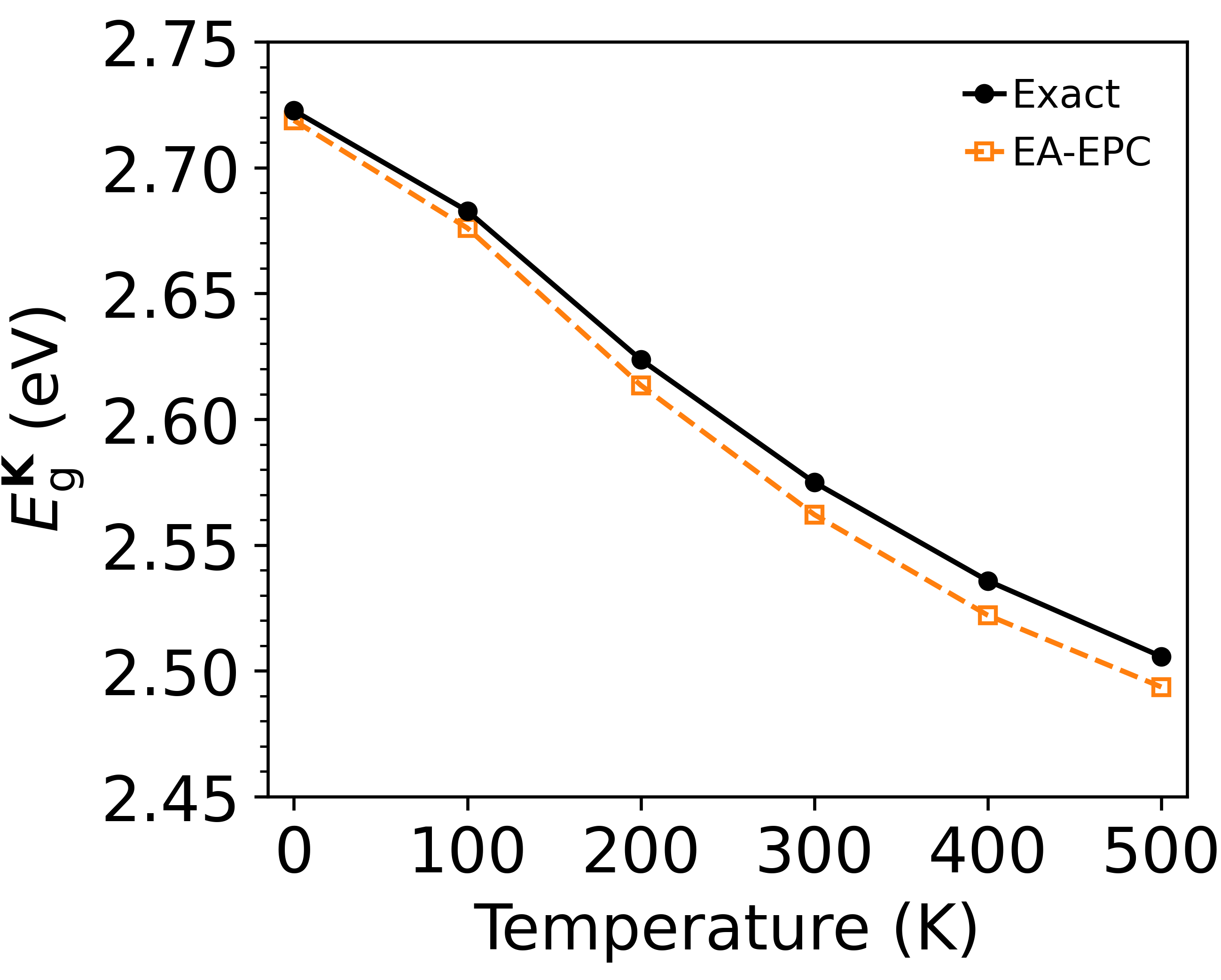}
\caption{Temperature dependence of the direct band gap at the K point of bilayer WSe$_2$. The exact calculation (black) is compared to the EA-EPC approach (orange).
\label{fig:wse2_bilayer_Eg_vs_T}}
\end{center}
\end{figure}

Figure~\ref{fig:wse2_bilayer_Eg_vs_T} shows the temperature dependence of the direct QP gap at K. The exact reference exhibits the expected gap closing, with a zero-point renormalization (ZPR) of 67 meV, obtained as the difference between the $G_0W_0$ gap of 2.790 eV and the $T=0$K gap value of 2.723 eV. The EA-EPC result reproduces the overall behavior of the exact curve across the entire $T=0$ K to 500 K window. The ZPR in this case amounts to 71 meV, a difference of only 4 meV. The EA-EPC result over-shoots slightly, but the deviation remains within a few meV.

\subsection{Pyridine/MoS$_2$}
\label{subsec:pyridine_mos2}

Having validated the addition step on the WSe$_2$ bilayer, we now assess the full approach on pyridine/MoS$_2$. As described in detail in Ref. \cite{GonzalezOliva2022,GonzalezOliva2026}, the MoS$_2$ unit cell is expanded into a 3$\times$3$\times$1 supercell when interfaced with a pyridine monolayer. Since the expansion step introduces no approximation (Eq.~\eqref{eq:expansion_dynmat}), it reproduces the exact supercell phonon dispersion; this is shown explicitly in Appendix~\ref{appendix:expansion_validation}.

\begin{figure}[t!]
\begin{center}
\includegraphics[width=0.95\columnwidth]{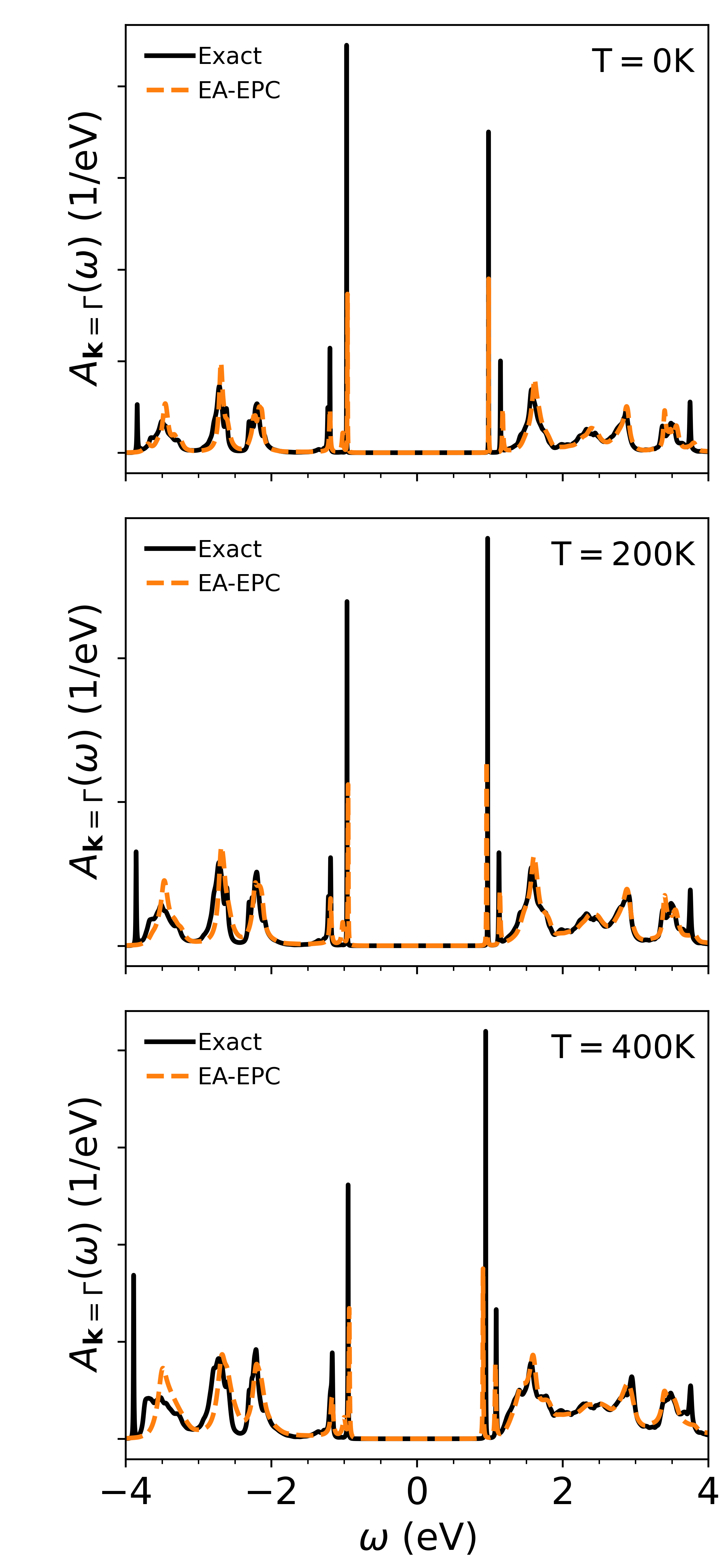}
\caption{Spectral functions of pyridine/MoS$_2$ at the $\Gamma$ point obtained with the EA-EPC approach (orange) and the exact calculation (black) for $T=0$K, $200$K, and $400$K.
\label{fig:spectral_function_hios}}
\end{center}
\end{figure}

\begin{figure}[t!]
\begin{center}
\includegraphics[width=0.95\columnwidth]{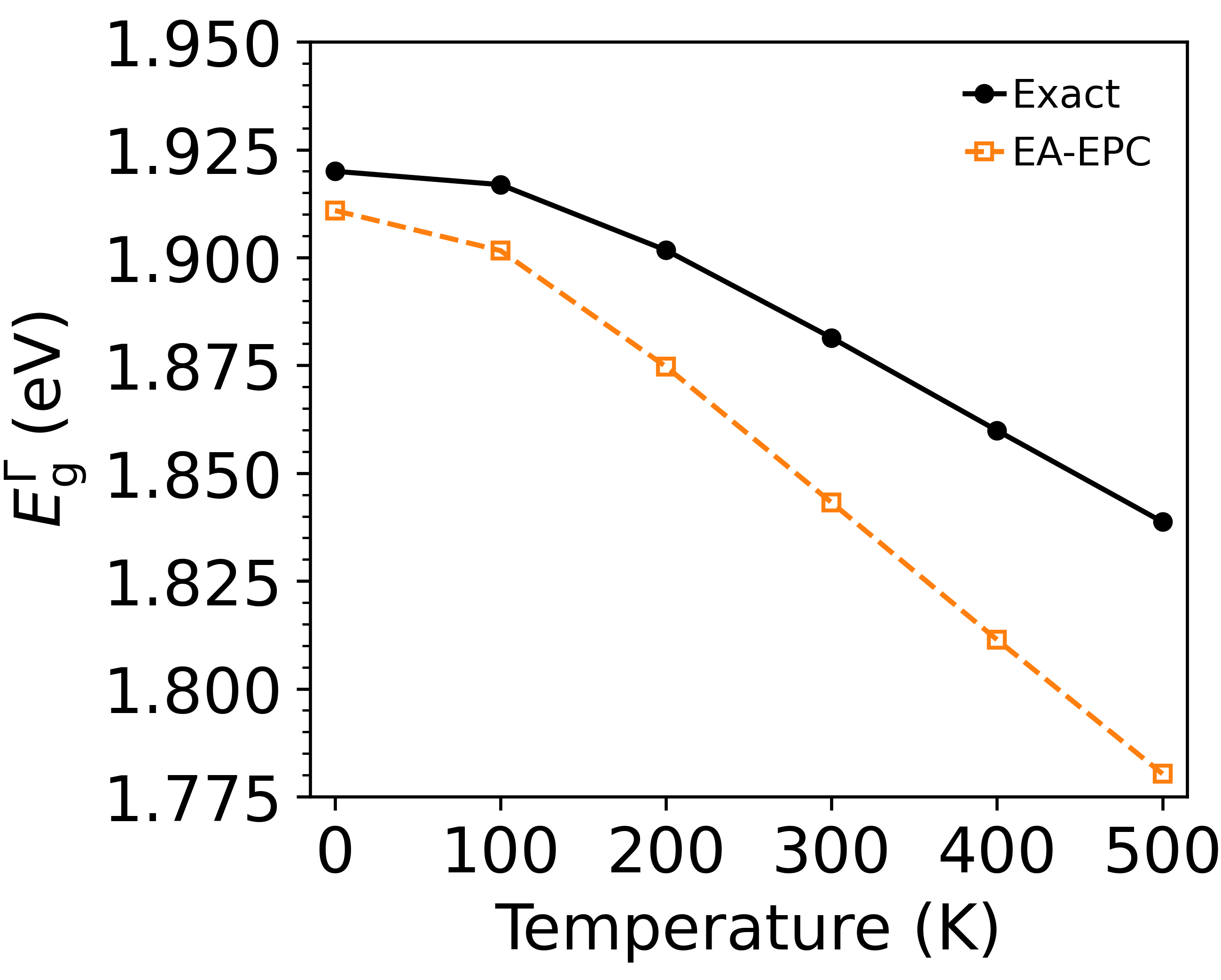}
\caption{Temperature dependence of the direct band gap at the $\Gamma$ point of pyridine/MoS$_2$. The exact supercell calculation (black) is compared to the EA-EPC approach (orange).
\label{fig:pyridine_mos2_Eg_vs_T}}
\end{center}
\end{figure}

Figure~\ref{fig:spectral_function_hios} shows the spectral function of pyridine/MoS$_2$ at the $\Gamma$ point for temperatures $T=0$ K, $200$ K, and $400$ K. Note that the fundamental gap of the supercell is located at this point due to the BZ folding of the MoS$_2$ bands. The EA-EPC approach (orange) reproduces the renormalization of the band-edge peaks of the exact calculation (black) across the full temperature range. The peak positions indicating the renormalized band gap as well as the temperature-induced shifts are overall well captured by the approximation. This is shown explicitly in Fig.~\ref{fig:pyridine_mos2_Eg_vs_T}. The EA-EPC approach yields a ZPR value of 90 meV, \ie 10 meV larger than the exact calculation. This slight overestimation of the EPC-induced band renormalization is also visible in the temperature dependence, where the two curves follow the same trend, and the deviation reaches 58 meV at 500 K. 

These results make obvious that our EA-EPC approach inherits the accuracy and limitations of the original EAS method~\cite{Xuan2019}: It is exact in the limit of zero interlayer coupling and becomes less accurate as the vdW character of the interface becomes less dominant. For the HIOS, this drawback is more pronounced than for WSe$_2$ bilayer, and its origin lies in the molecular vibrational modes, which are considerably more sensitive to the interlayer environment than those of MoS$_2$. The consequences for the spectral function are twofold. The MoS$_2$-derived band edges at $\Gamma$, and hence the band-gap renormalization are governed by the substrate phonons and are therefore reproduced by the EA-EPC approach, as discussed in Section \ref{subsec:wse2}. However, the valence-region satellites reported in Ref.~\cite{GonzalezOliva2026} are not recovered. These satellites originate from coupling of the MoS$_2$ valence states to the molecular vibrations. The addition step, which ignores this interlayer coupling, cannot reproduce such hybrid satellites, for which a full heterostructure calculation remains necessary. We expect the approximation to perform considerably better for heterostructures with larger, more rigid adsorbates. For example, in Ref.~\cite{GonzalezOliva2026}, the satellite of pyrene/MoS$_2$ was found to be predominantly of molecular origin. Such molecular satellites depend less on the interlayer coupling and should be captured well by the EA-EPC approach. More generally, larger and more densely packed molecular layers are stiffer and less perturbed by the substrate, reducing the interlayer sensitivity that limits the approximation for the example chosen here. A detailed assessment of the approach therefore calls for systematic benchmarking over stacking, interlayer distance, and character of the adsorbed molecules. 

\begin{table*}[t!]
\caption{\label{table:computational_efficiency} Comparison of computing resources for pyridine/MoS$_2$ between the exact method and the EA-EPC approach. The MoS$_2$ ingredients are obtained from a single unit cell on a $6\times6\times1$ $\mathbf{q}$-grid, the pyridine ingredients from a supercell on a $2\times2\times1$ $\mathbf{q}$-grid (see Section \ref{subsec:comp_details}). The exact reference uses a $2\times2\times1$ $\mathbf{q}$-grid for the full supercell.}
\begin{ruledtabular}
\begin{tabular}{lcc}
\textbf{Task} & \textbf{CPU hours (EA-EPC)} & \textbf{CPU hours (exact)} \\
\hline
Dynamical matrix + potential response (full)        & --      & 150{,}073 \\
Dynamical matrix + potential response  (MoS$_2$ unit cell) & 615 & -- \\
Dynamical matrix + potential response (pyridine monolayer)          & 14{,}976 & -- \\
Dynamical matrices: addition step                    & 0 & -- \\
Potential response: addition step    & 12{,}313 & -- \\
\cline{1-3}
\textbf{Total}                                  & 27{,}904 & 150{,}073 \\
\end{tabular}
\end{ruledtabular}
\end{table*}

Let us now turn to the gain in computational efficiency brought by the EA-EPC method, which after all enables calculations of EPC-renormalization effects on the electronic structure of interface systems. Table~\ref{table:computational_efficiency} compares the CPU hours~\footnote{CPU hours are defined as the number of processors used multiplied by the number of wall-time hours needed for the calculation.} of the exact reference with that of the EA-EPC calculation for pyridine/MoS$_2$. The exact reference, \ie the full DFPT calculation of the dynamical matrix and the potential response, requires $150{,}073$ CPU hours. In the EA-EPC, the computing time of the MoS$_2$ unit-cell ingredients amounts to $615$ CPU hours. Compared to the direct supercell calculation of the MoS$_2$ dynamical matrix and potential response ($44{,}839$ CPU hours), the expansion step itself adds negligible cost. The pyridine ingredients are obtained from a calculation using the same lattice parameters as the supercell, consuming $14{,}976$ CPU hours. Having calculated the ingredients of the two individual layers, the addition step for the dynamical matrices is performed at no extra computational time. However, the assembling of the potential response requires $12{,}313$ CPU hours, dominated by the interstitial to MT transformation of Eq.~\eqref{eq:addition_pot_mt}. The total of the EA-EPC calculations, summing the MoS$_2$ unit cell, the (negligible) expansion step, the pyridine supercell, and the addition of the potential response, amounts to $27{,}904$ CPU hours. This is, compared to the $150{,}073$ CPU hours of the exact reference, an overall saving of roughly $81\%$.

We emphasize that this saving applies to the EPC calculations alone. These can be combined with the EAS used to obtain the underlying $G_0W_0$ QP energies, which for the same system reduces the cost of the polarizability step by more than $50\%$~\cite{Schebek2025_PRB}. Looking further ahead, if BSE calculations are required, the EAS approach reduces the polarizability cost by almost $70\%$~\cite{Schebek2025_PRB}, thus making a fully many-body description of large vdW heterostructures tractable at a fraction of the cost of the corresponding exact supercell calculations.

\section{Summary and Conclusions}
\label{sec:summary}

In summary, we have presented an efficient approach to EPC in vdW heterostructures, which constructs the two key ingredients of the electron-phonon matrix, the dynamical matrices and the potential response, from those of the isolated components. This is achieved through an exact Fourier expansion from the unit cells onto the supercell followed by a blockwise addition that neglects (the weak) interlayer coupling. We have implemented this expansion and addition approach to EPC in the all-electron full-potential code \exciting. For the WSe$_2$ bilayer, both the phonon dispersion and the spectral function of the exact supercell reference are excellently reproduced. For pyridine/MoS$_2$, the method captures the band-gap renormalization and its temperature dependence, which are governed by the substrate, but not the hybrid molecular satellites, whose interlayer character lies beyond the approximation used. We therefore expect the approach to be most accurate for heterostructures with rigid, weakly coupled adsorbates. Combined with the EAS method, which constructs the non-interacting polarizability entering $G_0W_0$ and BSE calculations, our approach presented here, completes a unified MBPT framework for vdW heterostructures. This opens the door to systematic studies of temperature-dependent band structures, carrier dynamics, and phonon-assisted spectra in hybrid materials and other weakly interacting heterostructures, which are presently out of reach of direct calculations.

\section*{Acknowledgments}
This work was supported by the DFG (Deutsche Forschungsgemeinschaft) within the Collaborative Research Center HIOS (SFB 951), project 182087777. Partial support was also provided by the European Union’s Horizon 2020 Research and Innovation Program under the grant agreement no. 951786 (NOMAD CoE). We gratefully acknowledge the computing time on the high-performance computers ``Lise" and ``Emmy" at the NHR centers NHR@ZIB and NHR@Göttingen, and at the Paderborn Center for Parallel Computing (PC2). The NHR centers are jointly supported by the Federal Ministry of Education and Research and the state governments participating in the NHR, \url{www.nhr-verein.de}.

\section*{Data availability}
Input and output files of all calculations can be downloaded free of charge from the NOMAD data infrastructure~\cite{Draxl2019,Scheidgen2023} by the following link: \url{https://doi.org/10.17172/nomad.832d-df7c}.\\

\appendix

\section{Derivation of the expansion step}
\label{appendix:expansion}

\subsection{Dynamical matrix}

The supercell dynamical matrix follows from the interatomic force constants as
\begin{widetext}
\begin{align}
  D^{\rm sc}_{\alpha^{\rm sc} i,\alpha'^{\rm sc} j}(\mathbf{q}^{\rm sc})
  &= \sum_{\mathbf{R}^{\rm sc}}
     e^{-i\mathbf{q}^{\rm sc}\cdot\mathbf{R}^{\rm sc}}\,
     \Phi_{\alpha^{\rm sc} i,\alpha'^{\rm sc} j}(\mathbf{R}^{\rm sc})
     \nonumber \\[5pt]
  &= \sum_{\mathbf{R}^{\rm sc}}
     e^{-i\mathbf{q}^{\rm sc}\cdot\mathbf{R}^{\rm sc}}\,
     \Phi_{\alpha^{\rm uc} i,\alpha'^{\rm uc} j}
       \big(\mathbf{R}^{\rm sc}+\mathbf{R}^{\rm uc}_{\alpha'}-\mathbf{R}^{\rm uc}_{\alpha}\big)
     \nonumber \\[5pt]
  &= \sum_{\mathbf{R}^{\rm sc}}
     e^{-i\mathbf{q}^{\rm sc}\cdot\mathbf{R}^{\rm sc}}\,
     \frac{1}{N^{\rm uc}_{\mathbf{q}}}
     \sum_{\mathbf{q}^{\rm uc}}
     e^{i\mathbf{q}^{\rm uc}\cdot(\mathbf{R}^{\rm sc}+\mathbf{R}^{\rm uc}_{\alpha'}-\mathbf{R}^{\rm uc}_{\alpha})}\,
     D^{\rm uc}_{\alpha^{\rm uc} i,\alpha'^{\rm uc} j}(\mathbf{q}^{\rm uc})
     \nonumber \\[5pt]
  &= \frac{1}{N^{\rm uc}_{\mathbf{q}}}
     \sum_{\mathbf{q}^{\rm uc}}
     \Bigg[\sum_{\mathbf{R}^{\rm sc}}
       e^{i(\mathbf{q}^{\rm uc}-\mathbf{q}^{\rm sc})\cdot\mathbf{R}^{\rm sc}}\Bigg]\,
     e^{i\mathbf{q}^{\rm uc}\cdot(\mathbf{R}^{\rm uc}_{\alpha'}-\mathbf{R}^{\rm uc}_{\alpha})}\,
     D^{\rm uc}_{\alpha^{\rm uc} i,\alpha'^{\rm uc} j}(\mathbf{q}^{\rm uc}) \;,
     \label{eq:sm_dyn_expansion}
\end{align}
\end{widetext}
where $N_{\mathbf{q}}$ is the number of $\mathbf{q}$-points and $\mathbf{R}$ runs over the direct lattice vectors within the Born-von K\'arm\'an (BvK) cell. For commensurate $\mathbf{q}$-grids, unit cell and supercell share the same BvK cell, which contains $N^{\rm sc}_{\mathbf{q}}$ supercell lattice vectors. The lattice sum then becomes
\begin{equation}
  \sum_{\mathbf{R}^{\rm sc}}
  e^{i(\mathbf{q}^{\rm uc}-\mathbf{q}^{\rm sc})\cdot\mathbf{R}^{\rm sc}}
  = N^{\rm sc}_{\mathbf{q}}\,
    \delta_{\mathbf{q}^{\rm uc}\in\mathcal{F}(\mathbf{q}^{\rm sc})}\;,
  \label{eq:sm_lattice_sum}
\end{equation}
which is nonzero only for the set $\mathcal{F}(\mathbf{q}^{\rm sc})$ of unit-cell wavevectors that fold onto $\mathbf{q}^{\rm sc}$, \ie $\mathbf{q}^{\rm uc}=\mathbf{q}^{\rm sc}+\mathbf{G}^{\rm sc} (\mathbf{q}^{\rm uc})$, together with the equivalence of supercell and unit-cell atoms,
\begin{align}
  \boldsymbol{\tau}_{\alpha^{\rm sc}}
  & = \boldsymbol{\tau}_{\alpha^{\rm uc}} + \mathbf{R}^{\rm uc}_{\alpha}\;,
  \nonumber \\[5pt]
  \mathbf{R}^{\rm uc}_{\alpha'}-\mathbf{R}^{\rm uc}_{\alpha}
  & = (\boldsymbol{\tau}_{\alpha'^{\rm sc}}-\boldsymbol{\tau}_{\alpha'^{\rm uc}})
  - (\boldsymbol{\tau}_{\alpha^{\rm sc}}-\boldsymbol{\tau}_{\alpha^{\rm uc}})\;,
  \nonumber
\end{align}
one recovers the expansion of the dynamical matrix, Eq.~\eqref{eq:expansion_dynmat}.

\subsection{Potential response}

Consider a supercell atom $\alpha^{\rm sc}$ equivalent to unit-cell atom $\alpha^{\rm uc}$, \ie $\boldsymbol{\tau}_{\alpha^{\rm sc}}=\boldsymbol{\tau}_{\alpha^{\rm uc}}+\mathbf{R}^{\rm uc}_{\alpha}$, where $\mathbf{R}^{\rm uc}_{\alpha}$ is the unit-cell lattice vector locating the supercell atom $\alpha^{\rm sc}$ relative to its unit-cell equivalent (\ie indexed by the supercell atom). We transform the unit-cell potential response to real space,
\begin{equation}
  v^{{\rm uc}, (\mathbf{R}^{\rm uc}\alpha^{\rm uc} i)}_{\rm KS}(\mathbf{r})
  = \frac{1}{N_{\mathbf{q}}^{\rm uc}}\sum_{\mathbf{q}^{\rm uc}}
    e^{-i\mathbf{q}^{\rm uc}\cdot\mathbf{R}^{\rm uc}}\,
    v^{{\rm uc}, (\mathbf{q}^{\rm uc}\alpha^{\rm uc} i)}_{\rm KS}(\mathbf{r})\;,
    \label{eq:sm_transform_potential}
\end{equation}
where $\mathbf{R}^{\rm uc}$ is a unit-cell lattice vector in the BvK supercell. Since the unit cell and the supercell share the same BvK, displacing atom $\alpha^{\rm sc}$ in the supercell at $\mathbf{R}^{\rm sc}$ is the same perturbation as displacing atom $\alpha^{\rm uc}$ in the unit cell at $\mathbf{R}^{\rm sc}+\mathbf{R}^{\rm uc}_{\alpha}$, such that
\begin{equation}
v^{ {\rm sc},(\mathbf{R}^{\rm sc}\alpha^{\rm sc} i)}_{\rm KS}(\mathbf{r})
  = v^{{\rm uc},(\mathbf{R}^{\rm sc}+\mathbf{R}^{\rm uc}_{\alpha}\,\alpha^{\rm uc} i)}_{\rm KS}(\mathbf{r})\;.
  \label{eq:sm_pot_bvk_sc_uc}
\end{equation}
The inverse Fourier transform in the supercell then gives
\begin{align}
  v^{{\rm sc}, (\mathbf{q}^{\rm sc}\alpha^{\rm sc} i)}_{\rm KS}(\mathbf{r})
  & = \sum_{\mathbf{R}^{\rm sc}}
     e^{i\mathbf{q}^{\rm sc}\cdot\mathbf{R}^{\rm sc}}\,
     v^{{\rm sc},(\mathbf{R}^{\rm sc}\alpha^{\rm sc} i)}_{\rm KS}(\mathbf{r})
     \nonumber\\
& = \frac{1}{N^{\rm uc}_{\mathbf{q}}}\sum_{\mathbf{q}^{\rm uc}}\sum_{\mathbf{R}^{\rm sc}}
     e^{i(\mathbf{q}^{\rm sc}-\mathbf{q}^{\rm uc})\cdot\mathbf{R}^{\rm sc}}\,
     e^{-i\mathbf{q}^{\rm uc}\cdot\mathbf{R}^{\rm uc}_{\alpha}}\,
     \nonumber \\[5pt] & \times v^{{\rm uc},(\mathbf{q}^{\rm uc}\alpha^{\rm uc} i)}_{\rm KS}(\mathbf{r})\;.   
     \label{eq:sm_pot_expansion}
\end{align}
Using $\sum_{\mathbf{R}^{\rm sc}}e^{i(\mathbf{q}^{\rm sc}-\mathbf{q}^{\rm uc})\cdot\mathbf{R}^{\rm sc}}
=N^{\rm sc}_{\mathbf{q}}$ for all $\mathbf{q}^{\rm uc}$ with
$\mathbf{q}^{\rm uc}= \mathbf{q}^{\rm sc}+\mathbf{G}^{\rm sc} (\mathbf{q}^{\rm uc})$ and zero otherwise, one obtains the expansion of the potential response, Eq.~\eqref{eq:expansion_pot}.

\subsubsection{Muffin-tin contributions}

For $\mathbf{r}$ in the MT sphere of atom $\beta^{\rm sc}$
($\mathbf{r}=\boldsymbol{\tau}_{\beta^{\rm sc}}+\mathbf{s}$),
\begin{align}
  v^{{\rm sc},(\mathbf{q}^{\rm sc}\alpha^{\rm sc} i)}_{\rm KS}
    (\boldsymbol{\tau}_{\beta^{\rm sc}}+\mathbf{s})
  = & \frac{N^{\rm sc}_{\mathbf{q}}}{N^{\rm uc}_{\mathbf{q}}}
     \sum_{\mathbf{q}^{\rm uc}\in\mathcal{F}(\mathbf{q}^{\rm sc})}
     e^{-i\mathbf{q}^{\rm uc}\cdot\mathbf{R}^{\rm uc}_{\alpha}}\,
     \nonumber \\[5pt]
     & \times v^{{\rm uc},(\mathbf{q}^{\rm uc}\alpha^{\rm uc} i)}_{\rm KS}(\boldsymbol{\tau}_{\beta^{\rm uc}}+\mathbf{R}^{\rm uc}_{\beta}+\mathbf{s})\;,
\end{align}
and, using the $\mathbf{q}^{\rm uc}$ Bloch periodicity of $v^{{\rm uc},(\mathbf{q}^{\rm uc}\alpha^{\rm uc} i)}_{\rm KS}(\mathbf{r})$, one arrives at Eq.~\eqref{eq:expansion_pot_mt}.

\subsubsection{Interstitial contributions}

We seek the Fourier coefficients $\hat{v}^{(\mathbf{q}^{\rm sc}\alpha^{\rm sc} i)}_{\rm KS}(\mathbf{G}^{\rm sc})$ of the cell-periodic part $\Tilde{v}^{{\rm sc}, (\mathbf{q}^{\rm sc}\alpha^{\rm sc} i)}_{\rm KS}(\mathbf{r}) = e^{-i\mathbf{q}^{\rm sc}\cdot\mathbf{r}}\, v^{{\rm sc}, (\mathbf{q}^{\rm sc}\alpha^{\rm sc} i)}_{\rm KS}(\mathbf{r})$. From Eq.~\eqref{eq:expansion_pot}, we obtain:
\begin{align}
  \tilde v^{ {\rm sc},(\mathbf{q}^{\rm sc}\alpha^{\rm sc} i)}_{\rm KS}(\mathbf{r})
  &= \frac{N^{\rm sc}_{\mathbf{q}}}{N^{\rm uc}_{\mathbf{q}}}
     \sum_{\mathbf{q}^{\rm uc}\in\mathcal{F}(\mathbf{q}^{\rm sc})}
     e^{-i\mathbf{q}^{\rm uc}\cdot\mathbf{R}^{\rm uc}_{\alpha}}\,
     e^{i(\mathbf{q}^{\rm uc}-\mathbf{q}^{\rm sc})\cdot\mathbf{r}}\, \nonumber \\[5pt]
     & \times \tilde v^{{\rm uc}, (\mathbf{q}^{\rm uc}\alpha^{\rm uc} i)}_{\rm KS}(\mathbf{r})
     \nonumber\\[5pt]
  &= \frac{N^{\rm sc}_{\mathbf{q}}}{N^{\rm uc}_{\mathbf{q}}}
     \sum_{\mathbf{q}^{\rm uc}\in\mathcal{F}(\mathbf{q}^{\rm sc})}
     e^{-i\mathbf{q}^{\rm uc}\cdot\mathbf{R}^{\rm uc}_{\alpha}}\,
     e^{i\mathbf{G}^{\rm sc}(\mathbf{q}^{\rm uc})\cdot\mathbf{r}}
     \nonumber \\[5pt] & \times \sum_{\mathbf{G}^{\rm uc}}
     \hat{v}^{{\rm uc}(\mathbf{q}^{\rm uc}\alpha^{\rm uc} i)}_{\rm KS}(\mathbf{G}^{\rm uc})\,
     e^{i\mathbf{G}^{\rm uc}\cdot\mathbf{r}}\;,
     \label{eq:sm_pot_int_periodic}
\end{align}
where, in the second line, $\tilde{v}^{{\rm uc},(\mathbf{q}^{\rm uc}\alpha^{\rm uc} i)}_{\rm KS}$ is expanded in planewaves. Projecting Eq.~\eqref{eq:sm_pot_int_periodic} onto a supercell reciprocal-lattice vector $\mathbf{G}^{\rm sc}$ yields
\begin{align}
  \hat{v}^{{\rm sc}(\mathbf{q}^{\rm sc}\alpha^{\rm sc} i)}_{\rm KS}(\mathbf{G}^{\rm sc})
  &= \frac{1}{\Omega^{\rm sc}}
     \int_{\Omega^{\rm sc}}
     \tilde v^{{\rm sc},(\mathbf{q}^{\rm sc}\alpha^{\rm sc} i)}_{\rm KS}(\mathbf{r})\,
     e^{-i\mathbf{G}^{\rm sc}\cdot\mathbf{r}}\,\mathrm{d}\mathbf{r}
     \nonumber\\[5pt]
  &= \frac{N^{\rm sc}_{\mathbf{q}}}{N^{\rm uc}_{\mathbf{q}}}
     \sum_{\mathbf{q}^{\rm uc}\in\mathcal{F}(\mathbf{q}^{\rm sc})}
     e^{-i\mathbf{q}^{\rm uc}\cdot\mathbf{R}^{\rm uc}_{\alpha}}
     \nonumber \\[5pt] & \times \sum_{\mathbf{G}^{\rm uc}}
     \hat{v}^{{\rm uc}, (\mathbf{q}^{\rm uc}\alpha^{\rm uc} i)}_{\rm KS}(\mathbf{G}^{\rm uc})\, \nonumber \\[5pt] & \times \frac{1}{\Omega^{\rm sc}}\int_{\Omega^{\rm sc}}
     e^{i(\mathbf{G}^{\rm uc}+\mathbf{G}^{\rm sc}(\mathbf{q}^{\rm uc})-\mathbf{G}^{\rm sc})\cdot\mathbf{r}}
     \,\mathrm{d}\mathbf{r} \:.
     \label{eq:sm_pot_int_projection}
\end{align}
The integral enforces $\mathbf{G}^{\rm uc}=\mathbf{G}^{\rm sc}-\mathbf{G}^{\rm sc}(\mathbf{q}^{\rm uc})$, yielding the interstitial expansion, Eq.~\eqref{eq:expansion_pot_int}.

\section{Validation of the expansion step}
\label{appendix:expansion_validation}

\begin{figure}[t!]
\begin{center}
\includegraphics[width=0.95\columnwidth]{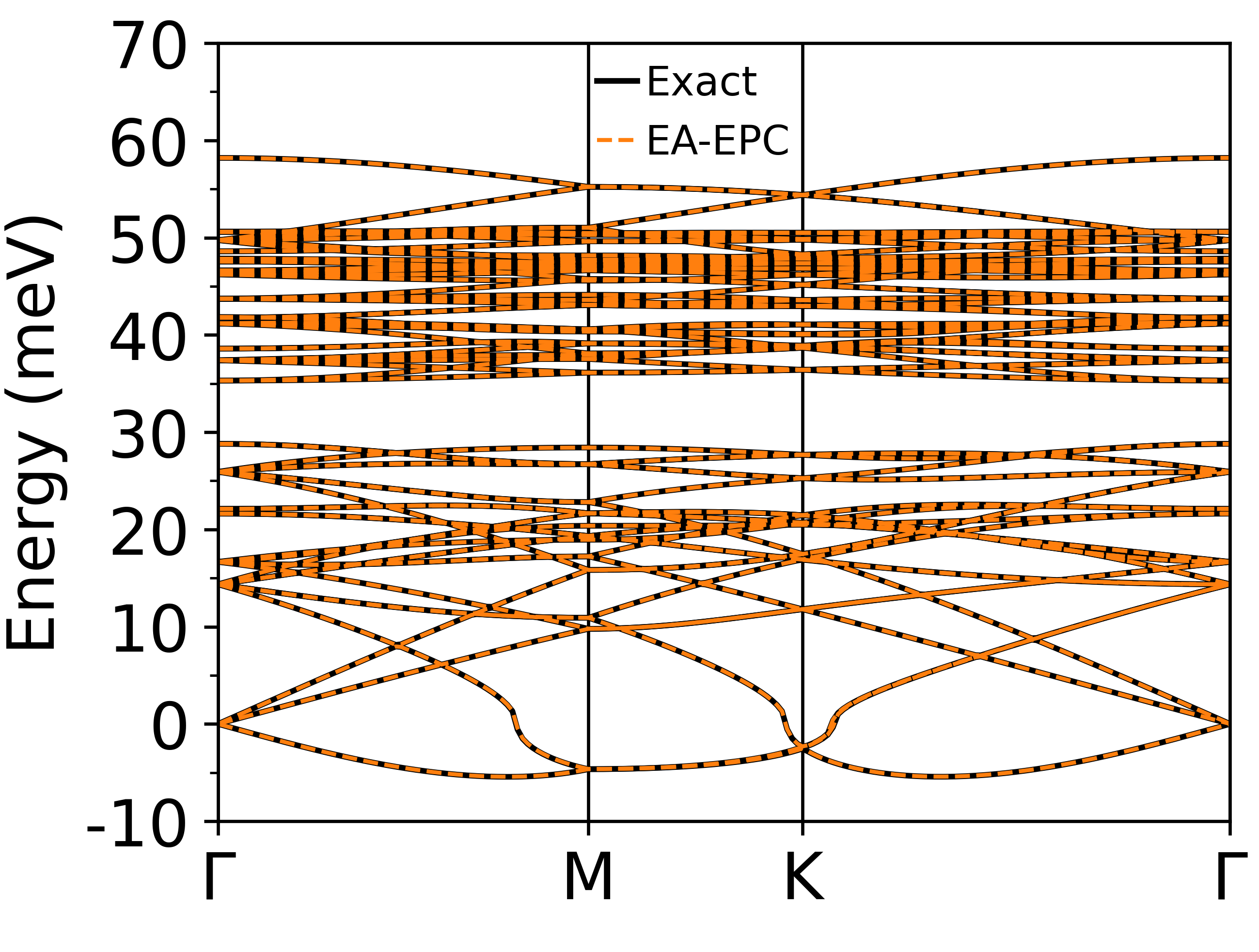}
\caption{Phonon dispersion of a $3\times3\times1$ supercell of MoS$_2$ obtained with the expansion step of the EA-EPC approach (orange) and with the exact calculation (black). 
\label{fig:phonon_dispersion_sc}}
\end{center}
\end{figure}

The phonon dispersion of the MoS$_2$ supercell obtained from the expansion step (Fig.~\ref{fig:phonon_dispersion_sc}, orange) perfectly reproduces the exact supercell calculation (black). This is expected, since, according to Eq.~\eqref{eq:expansion_dynmat}, the expansion does not introduce any approximation. 

\section{Derivation of the addition step}
\label{appendix:addition}

In the addition step, the heterostructure quantities are assembled from the (expanded) individual layers $\mu=1,2$. All layers and the heterostructure share the same supercell lattice and phonon $\mathbf{q}$-grid and, hence, the same real-space $\mathbf{R}$-set, so no Fourier transform or phase factor is required.

\subsection{Dynamical matrix}

Let $L(\alpha)$ denote the layer containing the heterostructure atom $\alpha$, and $\alpha\mapsto\alpha_L$ its index within layer $L$. For weakly bound systems, the interlayer force constants are neglected, so the heterostructure interatomic force constants reduce to a block copy of the intralayer contributions,
\begin{equation}
  \Phi^{\rm HS}_{\alpha i,\alpha' j}(\mathbf{R}) =
  \begin{cases}
    \Phi^{\mu}_{\alpha_\mu i,\alpha'_\mu j}(\mathbf{R}),
      & L(\alpha_\mu i) = L(\alpha'_\mu j) = \mu \\[5pt]
    0, & L(\alpha_\mu i) \neq L(\alpha'_\mu j) \:.
  \end{cases}
  \label{eq:app_addition_dyn_cases}
\end{equation}
Written compactly with Kronecker deltas, gives Eq.~\eqref{eq:addition_ifc}, \ie intralayer blocks are taken from the corresponding layer while all interlayer blocks are set to zero, reflecting the weak coupling.

\subsection{Potential response}

The heterostructure potential response is the sum of the layer responses,
\begin{equation}
  v^{{\rm HS}, (\mathbf{q}\alpha i)}_{\rm KS}(\mathbf{r})
  = v^{{\mu = 1},(\mathbf{q}\alpha i)}_{{\rm KS}}(\mathbf{r})
  + v^{{\mu = 2},(\mathbf{q}\alpha i)}_{{\rm KS}}(\mathbf{r})\;.
\end{equation}
Setting the response of the layer not containing the displaced atom $\alpha$ as zero, yields Eq.~\eqref{eq:addition_pot}.

\subsubsection{Muffin-tin region}

Let $\mathbf{r}$ be in the MT of atom $\beta$. If $\beta$ belongs to the same layer as the displaced atom $\alpha$, the MT representation of $v^{{L(\alpha)},(\mathbf{q}\alpha i)}_{{\rm KS}} (\boldsymbol{\tau}_\beta+\mathbf{s})$ is used directly. Otherwise $\beta$ lies in the interstitial of $L(\alpha)$, and the response is transformed from the interstitial to the MT representation,
\begin{align}
  v^{{\rm HS}, (\mathbf{q}\alpha i)}_{\rm KS}(\boldsymbol{\tau}_\beta+\mathbf{s})
  = \sum_{\mathbf{G}}
     \hat{v}^{{L(\alpha)},(\mathbf{q}\alpha i)}_{{\rm KS}}(\mathbf{G})\,
     e^{i(\mathbf{G}+\mathbf{q})\cdot(\boldsymbol{\tau}_\beta+\mathbf{s})} \:.
     \nonumber
\end{align}
Applying the Rayleigh expansion of the planewave $e^{i\mathbf{k}\cdot\mathbf{s}} = 4\pi\sum_{l,m} i^{l} j_l(|\mathbf{k}|s)\, Y^\ast_{lm}(\hat{\mathbf{k}})\, Y_{lm}(\hat{\mathbf{s}})$ with $\mathbf{k}=\mathbf{G}+\mathbf{q}$, one obtains Eq.~\eqref{eq:addition_pot_mt}.

\subsubsection{Interstitial region}

In the interstitial, the cell-periodic parts of the layer responses are taken directly from the layer containing the displaced atom, Eq.~\eqref{eq:addition_pot_int}. The interstitial representation always spans the full cell. The muffin-tins are removed only later, at the stage of evaluating the matrix elements, Eq.~\eqref{eq:eph_matrix_elements}, so no characteristic-function is applied during assembly.

\bibliography{bib.bib}

\end{document}